\documentclass[a4paper]{article}

\usepackage{18lomcon}        
\usepackage{cite}             
\usepackage{epsfig}           

\begin{document}


\title{HADRONIC RESONANCE PRODUCTION WITH ALICE AT THE LHC}

\author{Sergey Kiselev \email{Sergey.Kiselev@cern.ch}
        for the ALICE collaboration
}

\affiliation{Institute for Theoretical and Experimental Physics, 117259
Moscow, Russia}


\date{}
\maketitle


\begin{abstract}
We present recent results on short-lived hadronic resonances obtained
by the ALICE experiment in pp, p-Pb and Pb-Pb collisions at LHC energies,
including the most recent measurements of $\Lambda(1520)$ and $\Xi(1530)^{0}$ resonances.
\end{abstract}

Hadronic resonance production plays an important role both in elementary proton-proton and in heavy-ion collisions.
In heavy-ion collisions, since the lifetimes of short-lived resonances are comparable with
the lifetime of the late hadronic phase, regeneration and rescattering effects become
important and resonance ratios to longer lived particles can be used to estimate
the time interval between the chemical and kinetic freeze-out~\cite{Timescale}.
The measurements in pp and p-Pb collisions constitute a reference for nuclear collisions
and provide information for tuning event generators inspired by Quantum Chromodynamics.

Recent results on short-lived mesonic $\mathrm{K}^{*}(892)^{0}$, $\phi(1020)$ and baryonic $\Lambda(1520)$, 
$\Xi(1530)^{0}$ resonances (hereafter $\mathrm{K}^{*0}$, $\phi$, $\Lambda^{*}$, $\Xi^{*0}$) 
obtained by the ALICE experiment are presented.
The $\mathrm{K}^{*0}$ and $\phi$ have been measured in pp collisions at $\sqrt{s}$ = 13 TeV and in Pb-Pb collisions at $\sqrt{s_{\rm NN}}$ = 5.02 TeV
(results for the $\mathrm{K}^{*0}$ and $\phi$ in pp at $\sqrt{s}$ = 7 TeV, p-Pb at $\sqrt{s_{\rm NN}}$~=~5.02~TeV
and Pb-Pb at $\sqrt{s_{\rm NN}}$ = 2.76 TeV published in~\cite{ALICEpp7}, ~\cite{ALICEpPb} and  ~\cite{{ALICEPbPb}, {ALICEPbPb-highPT}}, respectively).
The $\Lambda^{*}$ has been measured in pp collisions at $\sqrt{s}$ = 7 TeV, in p-Pb collisions at $\sqrt{s_{\rm NN}}$~=~5.02~TeV 
and in Pb-Pb collisions at $\sqrt{s_{\rm NN}}$ = 2.76 TeV. 
The $\Xi^{*0}$ has been measured in Pb-Pb collisions at $\sqrt{s_{\rm NN}}$~=~2.76~TeV (results for the $\Sigma^{*\pm}$ and $\Xi^{*0}$ in pp at $\sqrt{s}$ = 7 TeV and p-Pb at $\sqrt{s_{\rm NN}}$~=~5.02~TeV published in~\cite{ALICEppSigmaStar} and ~\cite{ALICEpPbSigmaStar}, respectively).

The  resonances  are  reconstructed  in  their  hadronic  decay  channels  and have very different lifetimes. 
\begin{table}[ht]
\begin{center}
\begin{tabular}{ c c c c c }
\hline
                    &$\mathrm{K}^{*0}$  &  $\phi$ &    $\Lambda^{*}$    &     $\Xi^{*0}$      \\
\hline
decay channel (B.R.)&$\mathrm{K}\pi (0.67) $&$\mathrm{K}\mathrm{K} (0.49) $&$p\mathrm{K} (0.22)$&$\Xi\pi (0.67)$\\
lifetime (fm/\it{c})& 4.2 & 46.2 & 12.6 & 21.7 \\ 
\hline
\end{tabular}
\end{center}
\end{table}
%

\begin{figure}[hbtp]
\begin{center}
\includegraphics[scale=0.285]{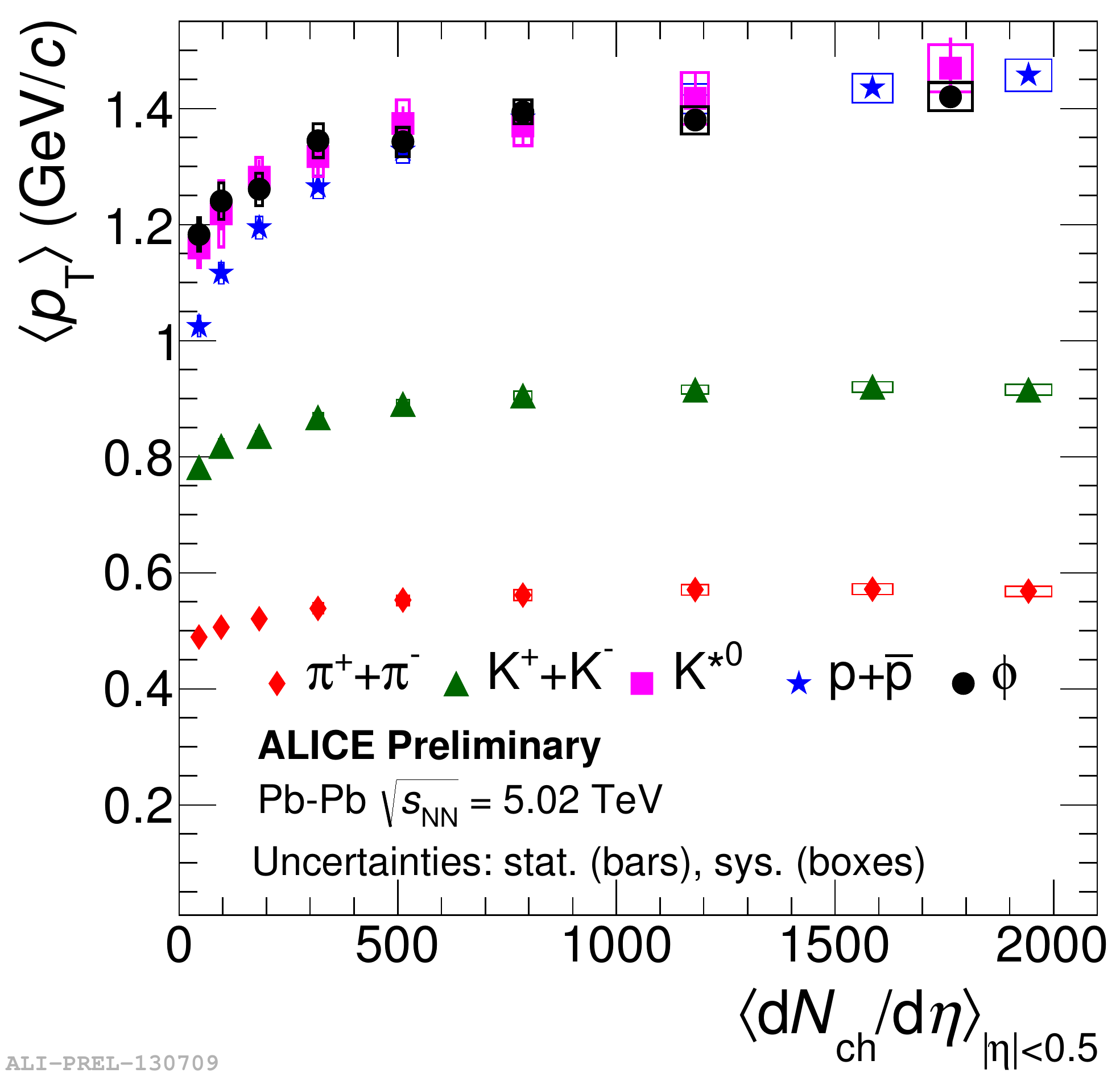}
\includegraphics[scale=0.295]{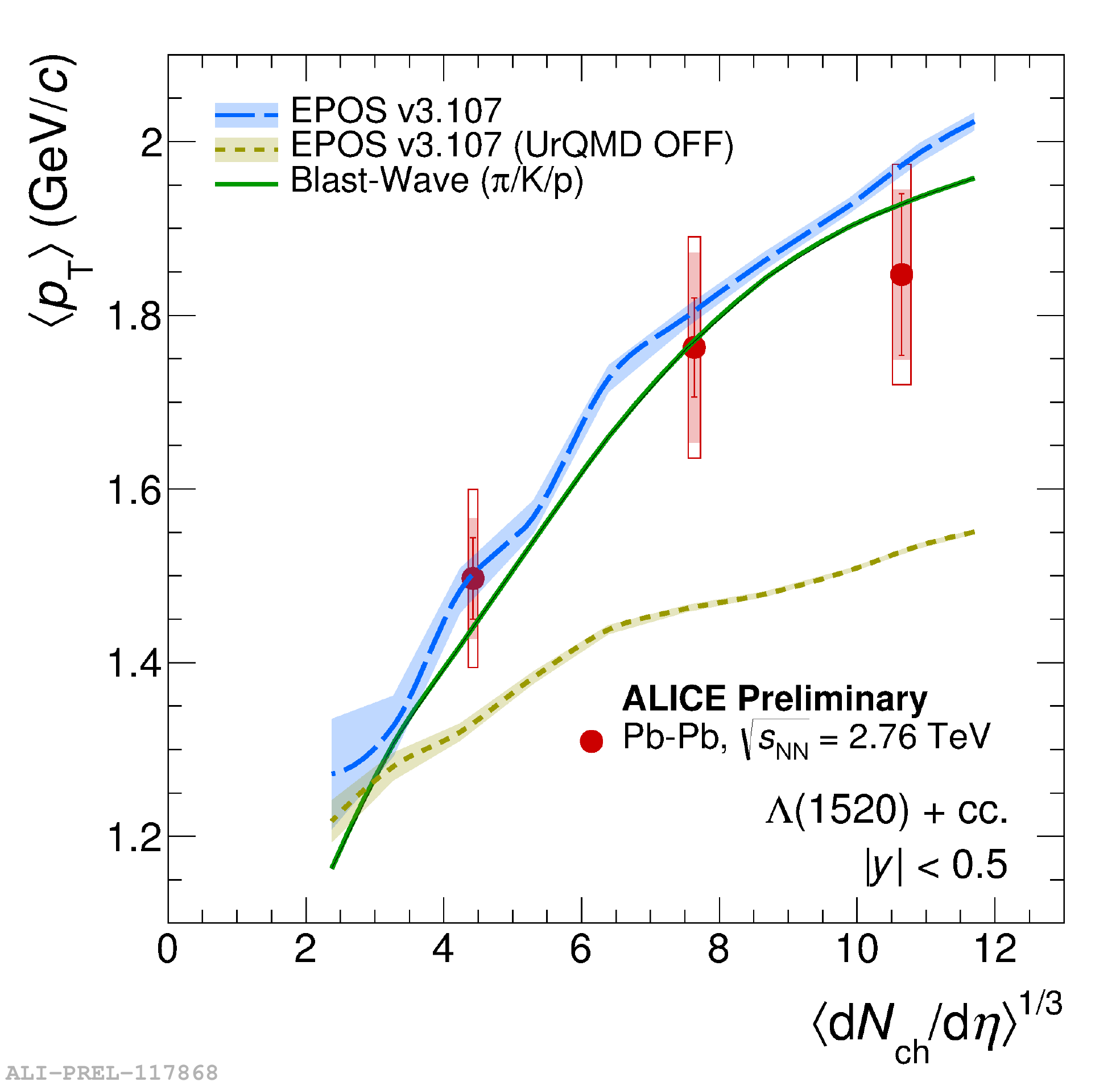}
\end{center}
\caption{(color online) 
(left) The mean transverse momentum of $\mathrm{K}^{*0}$, $\phi$ mesons and stable hadrons 
as a function of the mean charged particle multiplicity at mid-rapidity in Pb-Pb collisions at $\sqrt{s_{\rm NN}}$~=~5.02~TeV. 
(right) The mean transverse momentum of $\Lambda^{*}$ as a function of the mean charged particle multiplicity at mid-rapidity   
in Pb-Pb collisions at $\sqrt{s_{\rm NN}}$~=~2.76~TeV. The measurements are also compared to EPOS3~\cite{KnospeEPOS} and 
Blast-Wave~\cite{BlastWave} predictions.
}
  \label{fig:mpt}
\end{figure}

Figure~\ref{fig:mpt} (left) shows the mean transverse momentum $\langle p_\mathrm{T}\rangle$ of $\mathrm{K}^{*0}$, $\phi$ and stable hadrons
in Pb-Pb collisions at $\sqrt{s_{\rm NN}}$ = 5.02 TeV. 
The $\langle p_\mathrm{T}\rangle$ of resonances exhibits similar increasing trend with multiplicity as other hadrons. 
Moreover, mass ordering is observed in central Pb-Pb collisions. 
The $\mathrm{K}^{*0}$, p and $\phi$, which have similar masses, are observed to have
similar $\langle p_\mathrm{T}\rangle$ values, as expected if their spectral shape is dominated by radial flow. 
The results for Pb-Pb collisions at 5.02 TeV confirm the results in Pb-Pb collisions at 2.76 TeV~\cite{ALICEPbPb}.

Figure~\ref{fig:mpt} (right) presents the $\langle p_\mathrm{T}\rangle$ of $\Lambda^{*}$ in Pb-Pb collisions 
at $\sqrt{s_{\rm NN}}$ = 2.76 TeV. The results are in agreement with the prediction from the EPOS3 generator with 
UrQMD~\cite{KnospeEPOS}, which includes a modeling of re-scattering and regeneration in the hadronic phase.
The results are also in agreement with the average momentum extracted from the Blast-Wave model~\cite{BlastWave} 
with parameters obtained from the simultaneous fit to pion, kaon, and (anti)proton $p_{T}$ distributions~\cite{BlastWavePbPb}.

\begin{figure}[h]
\begin{center}
\includegraphics[scale=0.265]{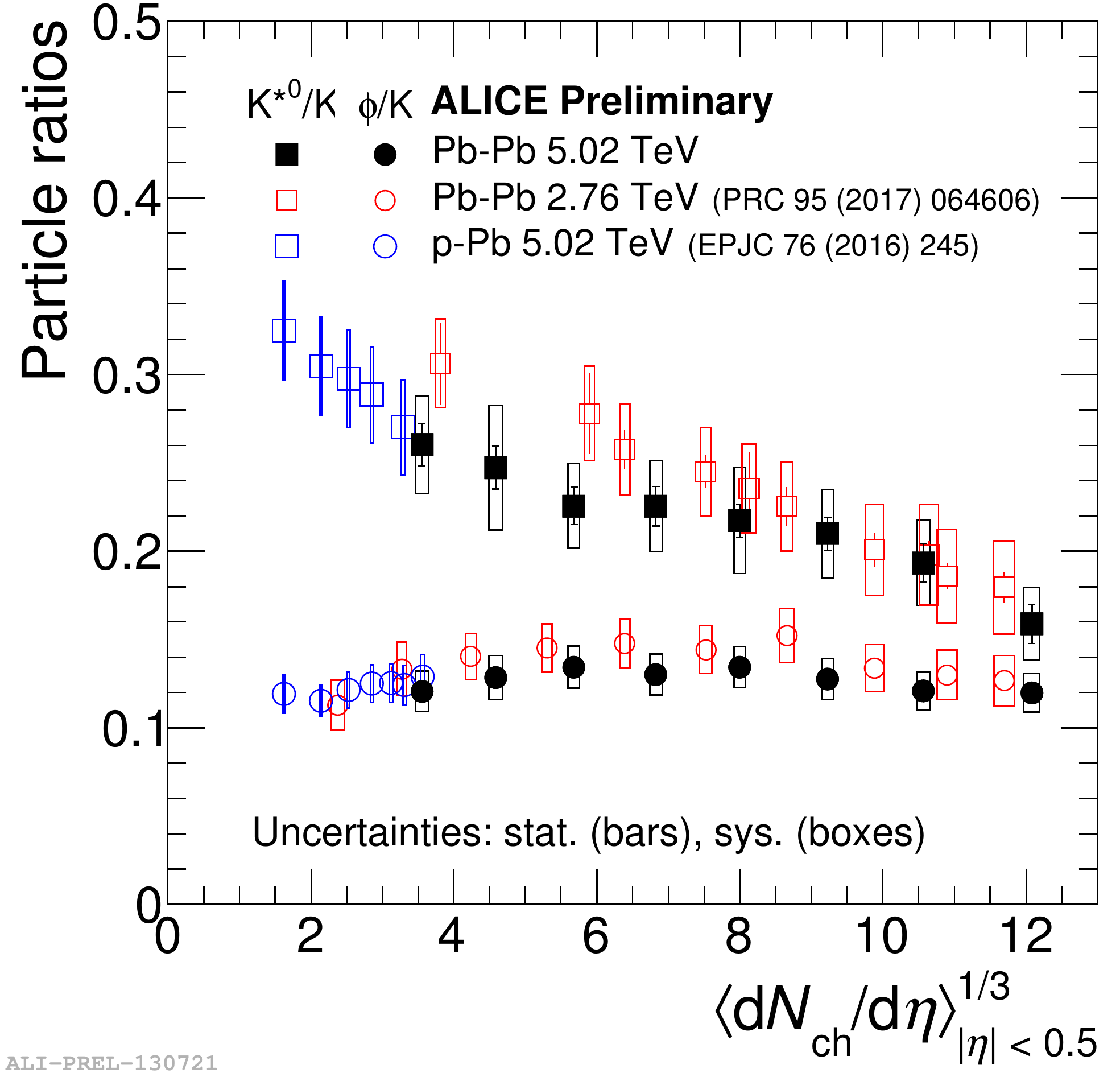}
\includegraphics[scale=0.315]{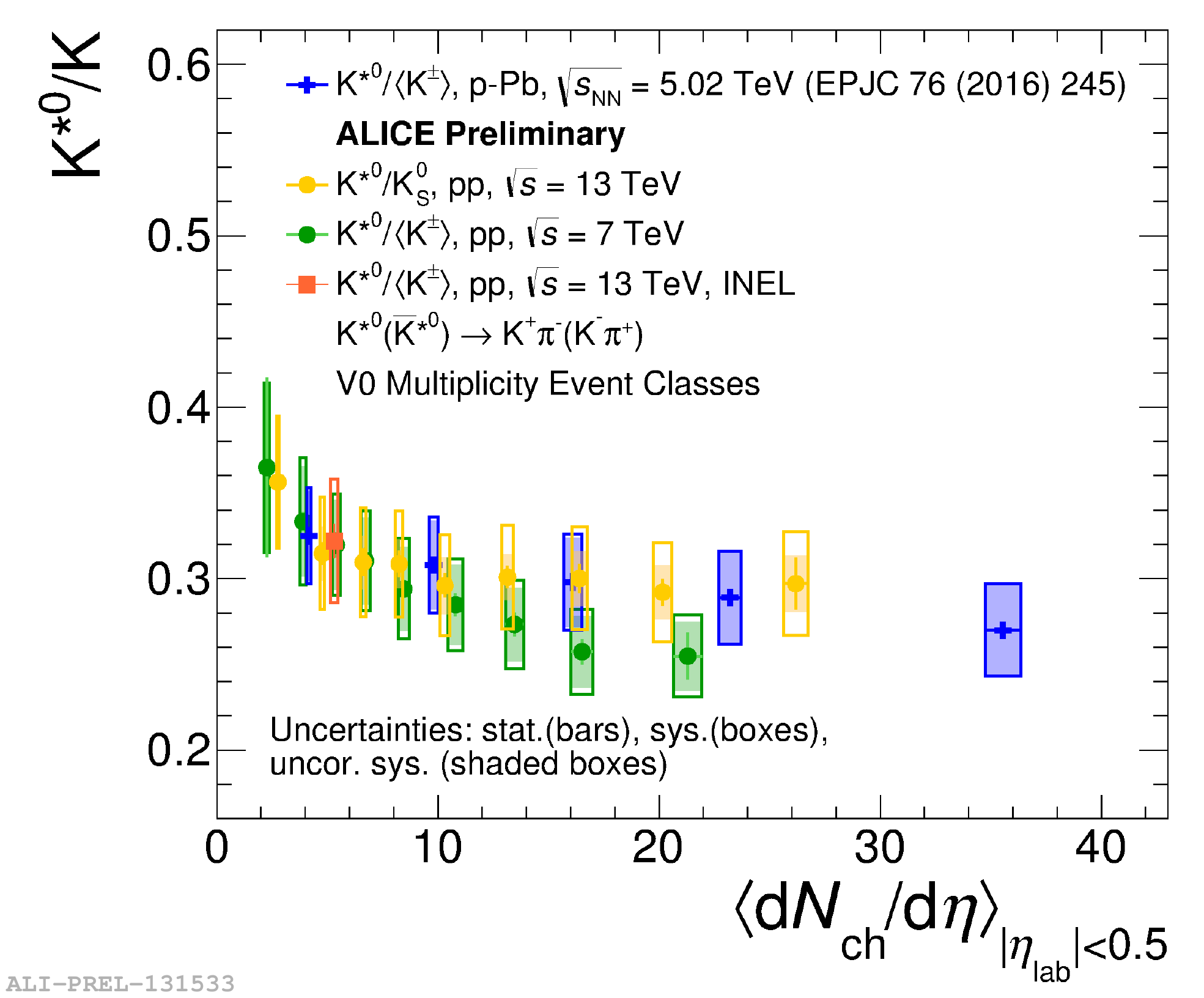}
\end{center}
\caption{(color online)  Particle yield ratios as a function of the mean charged particle multiplicity at mid-rapidity.
(left) $\mathrm{K}^{*0}/\mathrm{K}$ and $\phi/\mathrm{K}$ in Pb-Pb collisions at $\sqrt{s_{\rm NN}}$~=~5.02~TeV. 
Results for Pb-Pb  collisions at $\sqrt{s_{\rm NN}}$~=~2.76~TeV~\cite{ALICEPbPb-highPT} and p-Pb  collisions at $\sqrt{s_{\rm NN}}$~=~5.02~TeV~\cite{ALICEpPb} are also shown. 
(right) $\mathrm{K}^{*0}/\mathrm{K}$ in pp collisions at $\sqrt{s}$~=~13~TeV. Results for pp collisions at $\sqrt{s}$~=~7~TeV~\cite{ALICE-SQM16} 
and p-Pb collisions at $\sqrt{s_{\rm NN}}$~=~5.02~TeV~\cite{ALICEpPb} are also shown.  
}
  \label{fig:Mratios1}
\end{figure}

Figure~\ref{fig:Mratios1} (left) shows the particle yield ratios $\mathrm{K}^{*0}/\mathrm{K}$ and $\phi/\mathrm{K}$ 
in Pb-Pb collisions at $\sqrt{s_{\rm NN}}$~=~5.02~TeV. 
Results for Pb-Pb collisions at $\sqrt{s_{\rm NN}}$~=~2.76~TeV~\cite{ALICEPbPb-highPT} and p-Pb  collisions at $\sqrt{s_{\rm NN}}$~=~5.02~TeV~\cite{ALICEpPb} are also shown. 
The $\mathrm{K}^{*0}/\mathrm{K}$ ratio shows a significant suppression going from p-Pb and peripheral Pb-Pb collisions 
to most central Pb-Pb collisions. This suppression is consistent with rescattering of $\mathrm{K}^{*0}$ daughters in the hadronic 
phase of central collisions as the dominant effect and confirms the trend observed in Pb-Pb at 2.76 TeV~\cite{ALICEPbPb}.
The $\phi/\mathrm{K}$ ratio is nearly flat.
This suggests that rescattering effects are not important for $\phi$, which has 10 times longer lifetime 
than $\mathrm{K}^{*0}$ and decays mainly after the kinetic freeze-out.
 
Figure~\ref{fig:Mratios1} (right) presents the ratio $\mathrm{K}^{*0}/\mathrm{K}$ in pp collisions at $\sqrt{s}$~=~13~TeV. 
Results for pp at $\sqrt{s}$~=~7~TeV~\cite{ALICE-SQM16} and p-Pb at $\sqrt{s_{\rm NN}}$~=~5.02~TeV~\cite{ALICEpPb} are also shown.
There is a hint of decrease of the ratio with increasing multiplicity.
The values of the ratio are consistent for similar multiplicities across collision systems (pp, p-Pb) and energy (7, 13 TeV).
The decrease of the ratio might be an indication of a hadron-gas phase with non-zero lifetime in high-multiplicity pp and p-Pb collisions.

\begin{figure}[h]
\begin{center}
\includegraphics[scale=0.29]{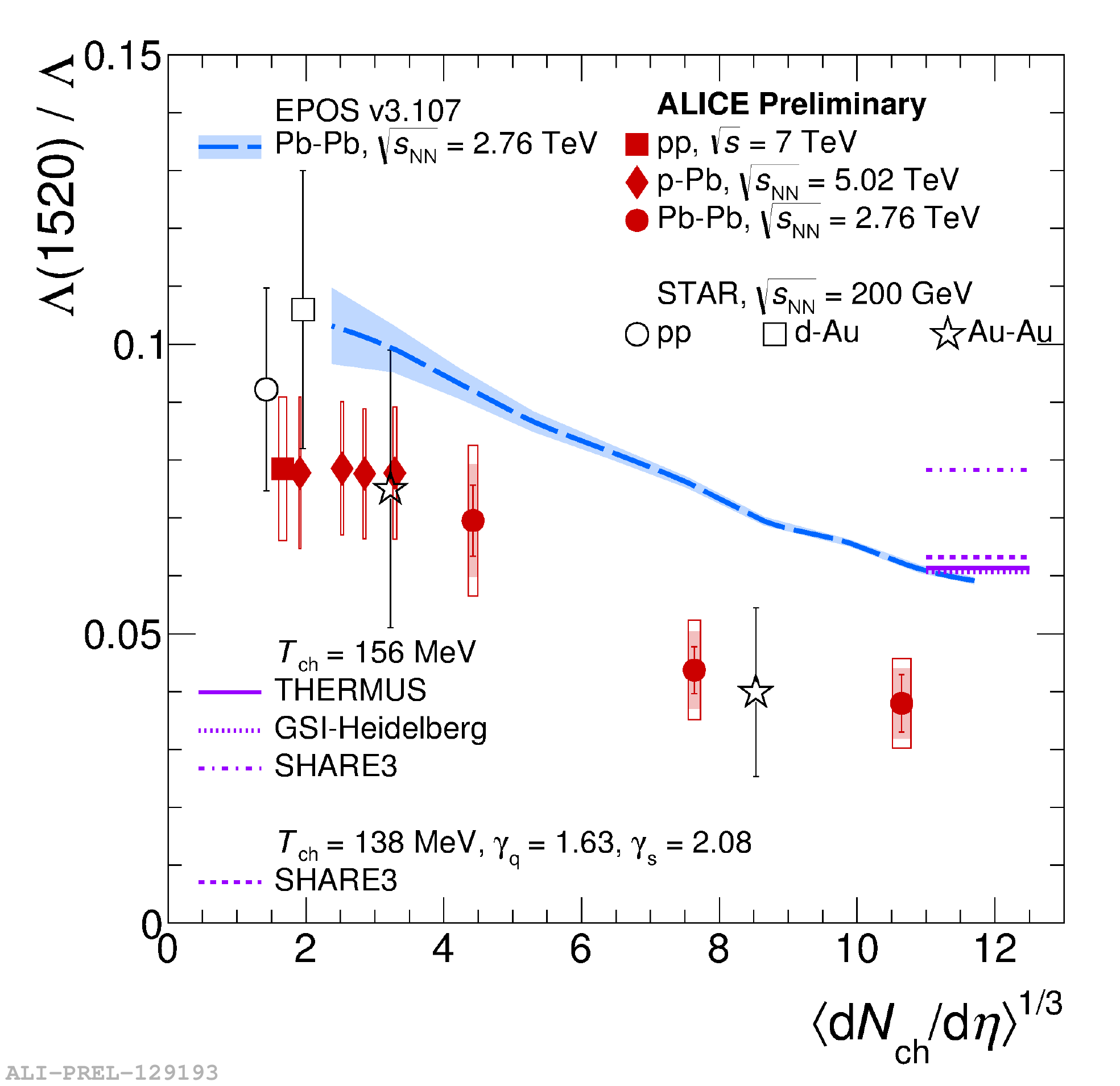}
\includegraphics[scale=0.29]{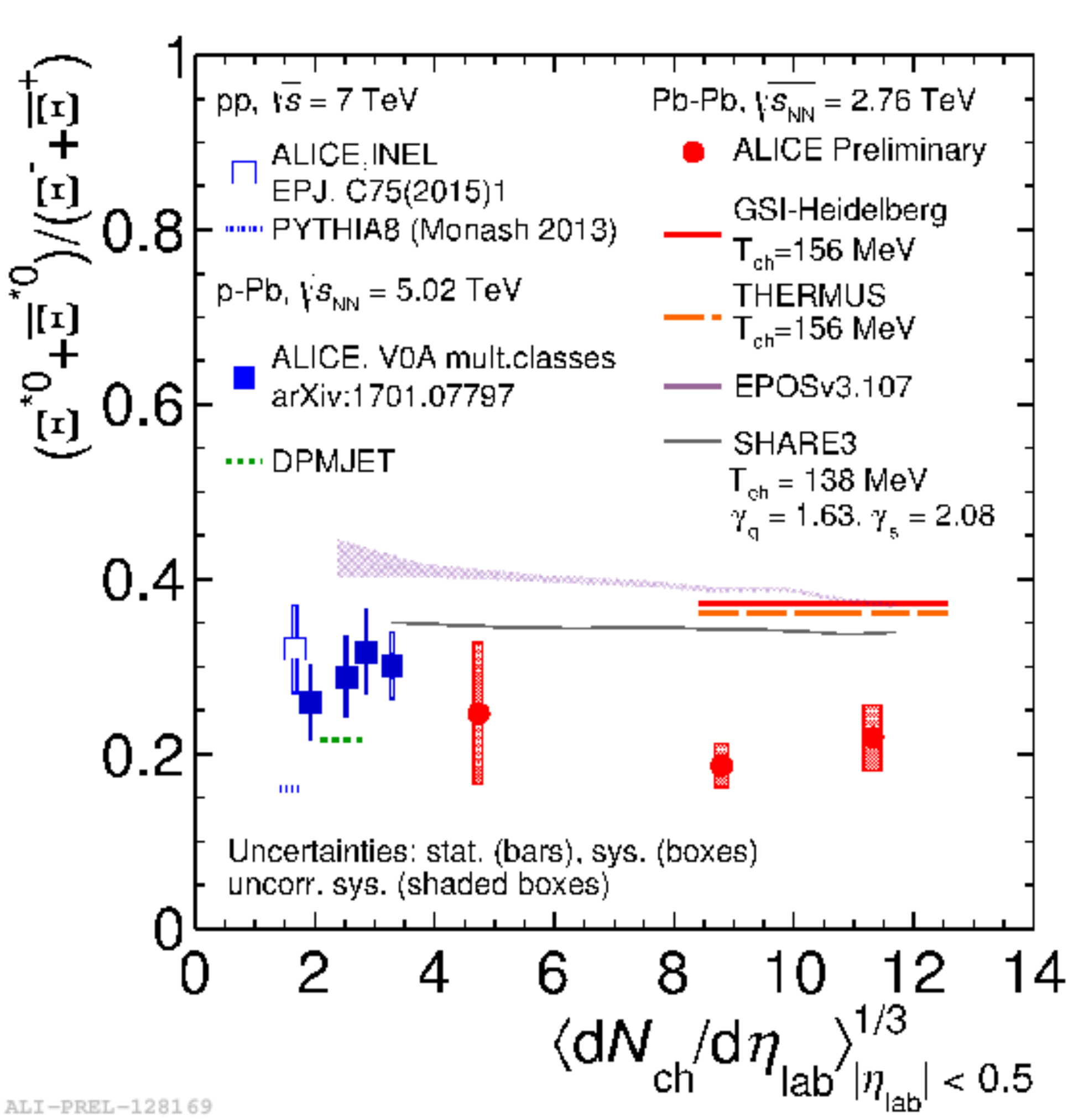}
\end{center}
\caption{(color online)  Particle yield ratios as a function of the mean charged particle multiplicity at mid-rapidity.
(left) $\Lambda^{*}/\Lambda$ for various collision systems. The measurements are also compared to model predictions: 
EPOS3~\cite{KnospeEPOS}, THERMUS~\cite{THERMUS}, GSI-Helderberg ~\cite{GSIHelderberg}, SHARE~\cite{SHARE}. STAR data from~\cite{STAR}. 
(bottom left)  $\Xi^{*0}/\Xi$ in Pb-Pb collisions at $\sqrt{s_{\rm NN}}$~=~2.76~TeV. Results for pp collisions at $\sqrt{s}$~=~7~TeV~\cite{ALICEppSigmaStar} 
and p-Pb collisions at $\sqrt{s_{\rm NN}}$~=~5.02~TeV~\cite{ALICEpPb} are also shown. 
The measurements are also compared to model predictions: PYTHIA8~\cite{PYTHIA8} and DPMJET~\cite{DPMJET}. 
}
  \label{fig:Mratios2}
\end{figure}

Figure~\ref{fig:Mratios2} (left) shows the particle yield ratio $\Lambda^{*}/\Lambda$ for various collision systems.
The $\Lambda^{*}/\Lambda$ ratio demonstrates a significant suppression going from pp, p-Pb and peripheral Pb-Pb collisions 
to most central Pb-Pb collisions. The suppression confirms the trend seen by STAR at $\sqrt{s_{\rm NN}}$~=~200~GeV~\cite{STAR}.
Although predictions of the EPOS3 model with UrQMD overestimate the data, the trend of the suppression is qualitatively reproduced.
In Pb-Pb collisions at $\sqrt{s_{\rm NN}}$~=~2.76~TeV the $\Lambda^{*}/\Lambda$ ratio suppression is similar to the behavior 
observed for the $\rho^{0}/\pi$~\cite{ALICE-ICPPA} and $\mathrm{K}^{*0}/\mathrm{K}$~\cite{ALICEPbPb-highPT} ratios.

Figure~\ref{fig:Mratios2} (right) presents the particle yield ratio $\Xi^{*0}/\Xi$ in Pb-Pb collisions at $\sqrt{s_{\rm NN}}$~=~2.76~TeV.
Results for pp collisions at $\sqrt{s}$~=~7~TeV~\cite{ALICEppSigmaStar} and p-Pb collisions at $\sqrt{s_{\rm NN}}$~=~5.02~TeV~\cite{ALICEpPb} are also shown. 
There is a hint of suppression in central Pb-Pb collisions with respect to pp and p-Pb collisions, but systematics are to be
reduced in peripheral Pb-Pb collisions before making any conclusive statement. EPOS3 with UrQMD overestimates the data and predicts only a slight decrease of the $\Xi^{*0}/\Xi$ ratio.

Thermal model predictions overestimate all particle ratios under study in central Pb-Pb collisions, except the $\phi/\mathrm{K}$ ratio.

\end{document}